\RequirePackage[displaymath]{lineno}
\documentclass[onecolumn,aps,prc,showpacs,superscriptaddress,floatfix,nofootinbib]{revtex4-2}
\usepackage{newtxtext,newtxmath,booktabs,siunitx}
\usepackage{dcolumn}
\usepackage{bm}
\usepackage{float}
\usepackage{ulem}
\usepackage[]{graphicx}  
\usepackage{booktabs}
\usepackage{tabularx}
\usepackage{amsmath}
\usepackage{xcolor}
\usepackage{multirow}
\usepackage[colorlinks,citecolor=blue,urlcolor=blue,linkcolor=blue]{hyperref}

\newcommand{\sigAA}{\sigma_{\rm AA}}
\newcommand{\signn}{\sigma_{\rm NN}}
\newcommand{\siggg}{\sigma_{\rm gg}}
\newcommand{\tpp}{T_{\rm pp}}
\newcommand{\snn}{\sqrt{s_{\rm NN}}}
\newcommand{\Pb}{${}^{208}$Pb}
\newcommand{\PbPb}{${}^{208}$Pb+${}^{208}$Pb}
\newcommand{\rnp}{\Delta r_{\rm np}}

\begin{document}

\title{Resolution-matched nuclear geometry and the nucleon-size ambiguity in relativistic heavy-ion collisions}

\author{Hao-jie Xu}
\email{haojiexu@zjhu.edu.cn}
\affiliation{School of Science and Strong-Coupling Physics International Research Laboratory (SPiRL), Huzhou Normal University, Huzhou, Zhejiang 313000, China}
\date{}

\begin{abstract}
Nuclear structure theory provides point-nucleon densities, whereas high-energy nuclear collisions probe nuclei through finite-resolution hadronic interactions.  This resolution mismatch becomes a physical ambiguity when point densities are embedded in Monte Carlo initial-state models with a finite transverse nucleon profile.  A parameter intended to describe the effective interaction range can then also reshape the nuclear surface, blurring the separation between nuclear structure and collision dynamics.  I show that this ambiguity can largely account for the strong nucleon-width dependence of the ${}^{208}$Pb+${}^{208}$Pb hadronic cross section ($\sigAA$) reported in recent Bayesian analyses.  Fixing the folded density that enters the Glauber phase shift removes this ambiguity at the level of nuclear geometry. The corrected cross section becomes nearly insensitive to a Gaussian nucleon width and instead probes the nuclear surface.
Within a two-component estimate for ${}^{208}$Pb, the current experimental uncertainty of $\sigAA$ translates into a broad neutron-skin interval, $\Delta r_{\rm np}\in[0,0.21]$ fm.
These results reframe $\sigAA$ as a surface-sensitive bridge between point-nucleon nuclear structure and finite-resolution high-energy initial conditions, rather than as a standalone nucleon-size observable.
This establishes resolution matching as a necessary step for using relativistic heavy-ion collisions as quantitative probes of nuclear structure.
\end{abstract}

\maketitle

\section{Introduction}
Relativistic heavy-ion collisions create a deconfined state of Quantum Chromodynamics (QCD) matter known as the quark--gluon plasma (QGP)~\cite{Adcox:2004mh,Adams:2005dq,ALICE:2010suc,Shuryak:2008eq}.  The collective flow of this matter is highly sensitive to the transverse geometry and fluctuations of the initial state~\cite{Ollitrault:1992bk,Kovtun:2004de,Romatschke:2007mq,Shen:2014vra,Xu:2016hmp,Zhao:2017yhj,Schenke:2020mbo,Bernhard:2016tnd,Bernhard:2019bmu,JETSCAPE:2020mzn,Nijs:2020roc,Gale:2013da,DerradideSouza:2015kpt,Noronha-Hostler:2015dbi,Song:2017wtw}.  This sensitivity has turned heavy-ion collisions into a new probe of nuclear structure~\cite{Li:2019kkh,Giacalone:2019pca,Jia:2021qyu,Xu:2021vpn,Zhang:2021kxj,Wang:2023yis,Xu:2024bdh,STAR:2024wgy,ALICE:2025luc,ATLAS:2025nnt,CMS:2025tga,STAR:2025ivi}.  Nuclear deformation, neutron skins, and cluster structures can all leave measurable imprints on final-state flow observables.  The program therefore begins with a simple but decisive input: the nuclear density used to initialize the collision.

Early heavy-ion simulations often used Woods--Saxon densities constrained by measured nuclear charge distributions~\cite{Miller:2007ri,Loizides:2014vua,Bernhard:2019bmu}.  This was a natural phenomenological starting point, but it does not determine the neutron density.  The missing proton--neutron asymmetry is a serious limitation for medium and heavy nuclei, and it can affect precision heavy-ion observables.  A well-known example is the predicted background in relativistic isobar collisions designed to search for the chiral magnetic effect~\cite{Xu:2017zcn,Li:2018oec,STAR:2019bjg}.

Recent calculations increasingly use nuclear-structure densities as initial-state input~\cite{Ryssens:2023fkv,Nijs:2022rme,Loizides:2016djv,Giacalone:2024luz}.  For medium and heavy nuclei, density-functional theory (DFT) provides a practical bridge between low-energy nuclear structure and event-by-event heavy-ion modeling~\cite{Chabanat:1997qh,Chamel:2009yx,Vautherin:1971aw,Baranger:1960qge,Bender:2003jk}.  It supplies separate proton and neutron density distributions and can encode neutron skins and deformation structure. Standard nuclear DFT, however, is formulated with point protons and point neutrons as active degrees of freedom.  The resulting densities are smooth one-body point-nucleon densities.  At this resolution, a nucleon-center density and a nuclear matter density are the same object: the point-nucleon density is the nuclear density.

In relativistic heavy ion collisions, Monte Carlo initial-state models add one more layer~\cite{Moreland:2014oya,dEnterria:2020dwq}. 
They assign a finite transverse profile to each sampled nucleon.
This profile represents subnucleonic, or effective interaction, structure at the resolution of a high-energy collision.
Once it is introduced, the sampled centers and the folded matter or interaction density are no longer identical. 
A point-nucleon density from nuclear structure theory must therefore be embedded carefully in the finite-width collision model.  This issue becomes even more serious in current Bayesian analyses, where this width is treated as a free parameter~\cite{Bernhard:2019bmu,Nijs:2020roc,JETSCAPE:2020mzn}.

This work addresses that problem through the hadronic nucleus--nucleus cross section ($\sigAA$).  The central issue is what physical density $\sigAA$ constrains when point-nucleon structure input is viewed through the finite resolution of a high-energy collision.  The Glauber framework makes the issue transparent because the cross section is controlled by the density entering the phase shift, especially near the nuclear surface.  I show that keeping the sampled point-center distribution fixed while varying the nucleon width changes this density and produces geometric surface inflation.  I then construct a density-preserving baseline in which the folded density, rather than the auxiliary center distribution, is held fixed.  This separates finite-kernel effects from changes of the nuclear geometry.  I apply the framework to the measured \PbPb\ cross section at $\snn=5.02$ TeV~\cite{ALICE:2022xir} and use it to infer the corresponding nuclear-surface constraint.

\section{Methods}
The Glauber framework describes high-energy nuclear cross sections in terms of the survival probability at a given impact parameter~\cite{Glauber:1959,Glauber:1970jm,Miller:2007ri,Loizides:2014vua,dEnterria:2020dwq}.  For a heavy nucleus, the interior is already close to opaque.  The total cross section is therefore governed mainly by the boundary region where the survival probability changes from nearly one to nearly zero.  This surface sensitivity makes hadronic cross sections an ideal place to define, and test, the density that enters the phase shift.

The starting point in many heavy-ion calculations is a Woods--Saxon density,
\begin{equation}
\rho^{\rm WS}(r)=\frac{\rho_0}{1+e^{(r-R)/a}},
\label{eq:ws}
\end{equation}
with radius $R$ and surface diffuseness $a$.  If the parameters are taken directly from measured charge densities, the profile already contains the finite electromagnetic size of the proton.  Using the same profile as a distribution of sampled nucleon centers, and then attaching a finite transverse profile to each nucleon, broadens the one-body density a second time.  This is geometric inflation of the nuclear surface.  Many calculations avoid this electromagnetic double counting by using point-nucleon radial distributions rather than the measured charge density itself~\cite{Loizides:2016djv,Nijs:2022rme,Giacalone:2023cet}.  That correction, however, does not exhaust the density-matching problem once the collision model also assigns a finite profile to each sampled nucleon.

The low-energy charge-density example is useful because it separates the nuclear density from the probe operator~\cite{Klos:2007is,Salcedo:1987md,Oset:1989ey}.  A DFT calculation gives point-proton and point-neutron densities.  The electromagnetic charge operator, including nucleon form factors and smaller relativistic corrections, is applied when comparing with electron scattering.  Charge radii and charge-density data can still enter the fitting of phenomenological functionals, but the finite-size correction then belongs to the model--data comparison rather than to an explicit finite-size nucleon degree of freedom.  Standard DFT therefore remains a point-nucleon theory at the level of its active densities.

The finite-width Glauber problem is the high-energy counterpart of this matching issue, but with one important difference.  The nucleon profile introduced in the Glauber calculation is not an external probe applied only after the nuclear density has been fixed.  Rather, it specifies the finite transverse resolution with which the point-nucleon geometry is sampled in the Glauber phase shift.  The corresponding width is therefore different from the electromagnetic charge radius; the charge-density example only illustrates the general need to match the nuclear density to the resolution kernel used in the observable.  In the point-nucleon limit, the DFT density can be used directly as the nuclear matter density entering the Glauber calculation.  Once a finite profile is attached to each sampled center, however, the Glauber phase shift depends on the folded density rather than on the center distribution.  A DFT or Woods--Saxon density used as nuclear input should therefore be matched to this folded density.  Using it instead as the center distribution makes the folded nuclear surface depend on the profile width, thereby mixing the finite interaction profile with the nuclear surface geometry.

This distinction is central to attempts to infer the nucleon width from the measured \PbPb\ cross section.  Ref.~\cite{Nijs:2022rme}, for example, used this cross section to constrain the nucleon width in the TRENTo framework.  Such an interpretation assumes that the change of $\sigAA$ with $w$ reflects the finite interaction kernel itself.  If the center distribution is kept fixed, varying $w$ changes two things at once: the kernel and the folded nuclear surface.  Treating $w$ as a free parameter in a Bayesian analysis then amplifies the ambiguity, because part of what is attributed to the nucleon width may instead be a change of the nuclear surface.  The construction below separates these effects by keeping the density entering the phase shift fixed.

The separation is most transparent at the level of the Glauber $S$ matrix.  In the TRENTo implementation~\cite{Moreland:2014oya}, the finite interaction range is specified by a nucleon width $w$ and by an effective parton--parton cross section $\siggg$.  The latter is not an independent observable.  For each chosen $w$, it is fixed by the measured inelastic nucleon--nucleon cross section,
\begin{equation}
\signn=\int d^2\mathbf b\left[1-\exp\left(-\siggg(w)\tpp(\mathbf b;w)\right)\right],
\label{eq:signn}
\end{equation}
with Gaussian overlap
\begin{equation}
\tpp(\mathbf b;w)=\frac{1}{4\pi w^2}\exp\left(-\frac{b^2}{4w^2}\right).
\label{eq:tpp_gauss}
\end{equation}
For \PbPb\ collisions at $\snn=5.02$ TeV I use $\signn=70$ mb.  Changing $w$ therefore changes the range of the $NN$ interaction and the value of $\siggg(w)$ needed to keep the measured $NN$ cross section fixed.

For a finite nucleon profile, the pairwise survival probability is
\begin{equation}
s_{NN}(\mathbf r;w)=\exp[-\siggg(w)\tpp(\mathbf r;w)],
\label{eq:sNN_pair}
\end{equation}
and the event-level nuclear $S$ matrix is
\begin{equation}
S_{AB}^{(e)}(\mathbf b;w)=\prod_{i\in A,j\in B}s_{NN}(\mathbf b+\mathbf s_i-\mathbf s_j;w),
\label{eq:S_MC_compact}
\end{equation}
where $\mathbf s_i$ are sampled transverse positions of nucleon centers.  The corresponding Monte Carlo cross section is
\begin{equation}
\sigma_{AB}^{\rm MC}(w)=\int d^2\mathbf b\left[1-\left\langle S_{AB}^{(e)}(\mathbf b;w)\right\rangle_{\rm events}\right].
\label{eq:sigma_MC_compact}
\end{equation}
In practice, I sample impact parameters uniformly in area up to $b_{\rm max}=20$ fm and evaluate
\(
\sigma_{AB}^{\rm MC}(w)=\pi b_{\rm max}^{2}\frac{N_{\rm interact}}{N_{\rm total}},
\label{eq:sigma_MC_counting}
\)
where $N_{\rm interact}$ counts events with at least one inelastic nucleon--nucleon interaction.  Purely electromagnetic and diffractive-dissociation events are not included, consistent with the ALICE definition of the hadronic cross section.  The same estimator is used for the default and density-preserving calculations; only the sampled center distribution is changed.

The familiar optical expression $S_{AB}=\exp[-\signn T_{AB}^{\rm pt}]$ is recovered when the finite $NN$ interaction profile is collapsed into a zero-range, point-nucleon coarse graining.  At that resolution, the point thickness also represents the nuclear mass thickness, as in point-nucleon nuclear-structure calculations.  Once a finite profile is attached to the sampled centers, however, the point thickness is no longer the finite-width density entering Eq.~(\ref{eq:S_MC_compact}).

In the smooth-density limit, Eq.~(\ref{eq:S_MC_compact}) becomes
\begin{equation}
S_{AB}^{\rm opt,X}(\mathbf b;w)=\exp[-\siggg(w){\cal T}_{AB}^{\rm X}(\mathbf b;w)],
\label{eq:S_opt_compact}
\end{equation}
where ${\rm X}$ denotes the density prescription.  If one keeps the point-center density fixed while varying $w$, the overlap entering the phase shift is
\begin{align}
{\cal T}_{AB}^{\rm std}(\mathbf b;w)
&=\int d^2\mathbf s_A d^2\mathbf s_B\,
T_A^{\rm pt}(\mathbf s_A)T_B^{\rm pt}(\mathbf s_B)
\tpp(\mathbf b+\mathbf s_A-\mathbf s_B;w)\nonumber\\
&=\int d^2\mathbf s\,
T_A^{{\rm mass,std}}(\mathbf s;w)T_B^{{\rm mass,std}}(\mathbf s+\mathbf b;w).
\label{eq:TAB_std}
\end{align}
This equation exposes the source of the ambiguity.  A fixed distribution of point centers does not define a fixed finite-width nucleus when the nucleon profile has a finite width.  Instead, it generates a $w$-dependent folded mass thickness $T^{\rm mass,std}$.

For the optical-limit cross section
\begin{equation}
\sigma_{AB}^{\rm opt,X}(w)=\int d^2\mathbf b\,[1-S_{AB}^{\rm opt,X}(\mathbf b;w)],
\label{eq:sigma_optical}
\end{equation}
the derivative with respect to $w$ separates into two physically different terms,
\begin{align}
\frac{d\sigma_{AB}^{\rm opt,X}}{dw}
=&\frac{d\siggg}{dw}\int d^2\mathbf b\,
S_{AB}^{\rm opt,X}(\mathbf b;w){\cal T}_{AB}^{\rm X}(\mathbf b;w)\nonumber\\
&+\siggg(w)\int d^2\mathbf b\,
S_{AB}^{\rm opt,X}(\mathbf b;w)
\frac{\partial {\cal T}_{AB}^{\rm X}(\mathbf b;w)}{\partial w}.
\label{eq:slope_split}
\end{align}
The first term is a residual finite-kernel effect associated with fixing $\signn$ through Eq.~(\ref{eq:signn}).  The second term is a change of the nuclear mass overlap itself.  For the standard prescription, $\partial_w{\cal T}_{AB}^{\rm std}$ is generated by the explicit $w$ dependence of the finite $NN$ kernel at fixed point-center density.  In the corrected prescription, the input center distribution changes with $w$ so that the final mass overlap remains fixed; hence $\partial_w{\cal T}_{AB}^{\rm corr}\simeq0$ by construction.  For a large nucleus, this geometric term acts mainly through the surface.  The central overlap is already highly opaque, while the total cross section is set by the boundary where the survival probability changes rapidly.  The strong apparent width dependence of the standard prescription is therefore dominated by the second contribution, for which $\partial_w{\cal T}_{AB}^{\rm std}\ne0$.  The corrected baseline therefore removes the surface-inflation contribution by imposing
\begin{equation}
{\cal T}_{AB}^{\rm corr}(\mathbf b)=\int d^2\mathbf s\,T_A^{\rm mass}(\mathbf s)T_B^{\rm mass}(\mathbf s+\mathbf b),\qquad
\partial_w {\cal T}_{AB}^{\rm corr}\simeq0.
\label{eq:TAB_corr}
\end{equation}
In this construction the folded mass thickness, not the auxiliary point-center distribution, is the nuclear input to the Glauber phase shift.  The point-center distribution is then chosen only as a sampling antecedent that generates the desired density after convolution with the nucleon profile.  Thus the finite-width simulation is calibrated to the same nuclear density that would be used in the point-nucleon coarse graining, rather than to the uncorrected distribution of sampling centers.

The equations above define the fixed-density condition at the level of the transverse Glauber overlap.  For event generation, it is most convenient to impose the same condition on the three-dimensional density before projection to the transverse thickness.  Let $f(\bm\xi;w)$ be the distribution of nucleon centers and $K_p(\mathbf r,\bm\xi;w)$ the three-dimensional nucleon profile.  The generated physical mass density is
\begin{equation}
\rho(\mathbf r;w)=\int d^3\xi\, f(\bm\xi;w)K_p(\mathbf r,\bm\xi;w).
\label{eq:Weierstrass}
\end{equation}
For a Gaussian profile,
\begin{equation}
K_p(\mathbf r,\bm\xi;w)=\frac{1}{(2\pi w^2)^{3/2}}
\exp\left[-\frac{(\mathbf r-\bm\xi)^2}{2w^2}\right].
\label{eq:kernel}
\end{equation}
Equation~(\ref{eq:Weierstrass}) is a three-dimensional Weierstrass transform.  The density-preserving point-center distribution is therefore the inverse-transform antecedent of the target nuclear density.  This construction can be applied to either a Woods--Saxon target density or to proton and neutron densities from DFT.  Similar finite-size corrections appear in related transport contexts, such as quantum molecular dynamics~\cite{TMEP:2016tup,Yang:2021gwa,Xiang:2025yzm}; here the target is specifically the density entering the Glauber phase shift.  For the numerical demonstration below, I approximate the antecedent by a Woods--Saxon form with modified parameters $(\tilde\rho_0,\tilde R,\tilde a)$~\cite{Klos:2007is,Salcedo:1987md,Oset:1989ey,Xu:2021vpn}.

For each $w$, I determine the modified parameters by matching the first three radial moments of the convolved density to those of the target Woods--Saxon density,
\begin{align}
\langle \rho(r;w)\rangle &= \langle \rho^{\rm WS}(r)\rangle,\label{eq:norm}\\
\langle r\rho(r;w)\rangle &= \langle r\rho^{\rm WS}(r)\rangle,\label{eq:Wsfit}\\
\langle r^2\rho(r;w)\rangle &= \langle r^2\rho^{\rm WS}(r)\rangle,\label{eq:rms}
\end{align}
where $\langle r^n g\rangle\equiv\int r^n g(\mathbf r)d^3r/\int g(\mathbf r)d^3r$.

\begin{figure}[t]
\centering
\includegraphics[width=0.49\textwidth]{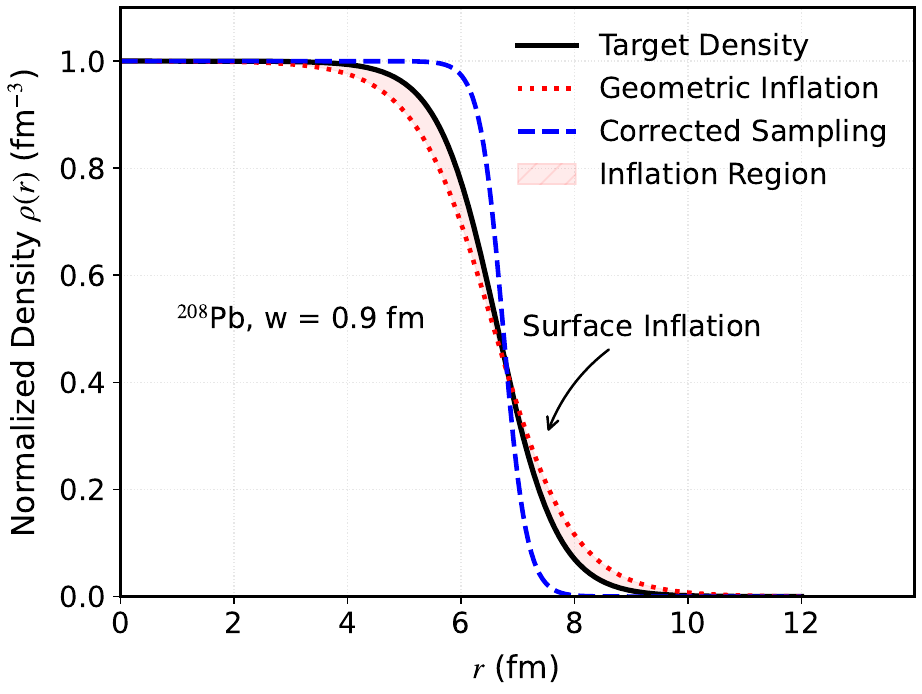}
\caption{Illustration of geometric inflation.  The target Woods--Saxon distribution is compared with the inflated density from uncorrected sampling for $w=0.9$ fm.  The corrected sampling distribution, obtained through an inverse-transform construction, ensures that the physical density matches the target after convolution.}
\label{fig:illustration}
\end{figure}

Figure~\ref{fig:illustration} shows the geometry of the correction.  If the target Woods--Saxon density is used directly as the center distribution, convolution with a finite nucleon profile broadens the resulting density.  The corrected center distribution is correspondingly sharper and should be understood as the auxiliary proton- or nucleon-center distribution needed to reproduce the same target density after convolution.

For \Pb, I use the conventional Woods--Saxon parameters $R=6.647$ fm and $a=0.523$ fm constrained by charge-density data~\cite{Fricke:1995zz} as a reference empirical surface.  This empirical profile provides a common target surface for demonstrating the density-preserving construction; the neutron-skin scan below then translates the same surface constraint into two-component proton and neutron matter profiles.  The density-preserving antecedent parameters can be estimated analytically by setting $\tilde\rho_0\equiv\rho_0$ and solving Eqs.~(\ref{eq:norm}) and (\ref{eq:rms})~\cite{Salcedo:1987md,Oset:1989ey}:
\begin{align}
\tilde R &\approx R+w^2\frac{15R}{15R^2+7\pi^2a^2},\nonumber\\
\tilde a &=\sqrt{a^2-\frac{\tilde R^3-R^3}{\pi^2\tilde R}}.
\label{eq:ana}
\end{align}
The leading behavior, $\tilde a^2\approx a^2-3w^2/\pi^2$, shows that a finite nucleon width must be compensated by a sharper center distribution if the final surface is to remain unchanged.  Equation~(\ref{eq:ana}) also reproduces the familiar finite-size correction used to obtain a proton-center, or point-proton, Woods--Saxon distribution.  As a check of the inverse-convolution formula, inserting an electromagnetic finite-size scale $w=0.51$ fm ($r_p=0.875$ fm) gives $\tilde R=6.682$ fm and $\tilde a=0.447$ fm, close to commonly used point-proton parameters~\cite{Loizides:2016djv,Nijs:2022rme,Giacalone:2023cet}.  In the present work, the same logic is applied to the finite-width mass density entering the Glauber phase shift.  The construction also illustrates that the inverse problem cannot be extended indefinitely: for \Pb, widths above roughly $w\simeq0.9$ fm would require an unphysical antecedent diffuseness, consistent with liquid-drop expectations~\cite{Luzum:2023gwy}.

\begin{figure}[t]
\centering
\includegraphics[width=0.49\textwidth]{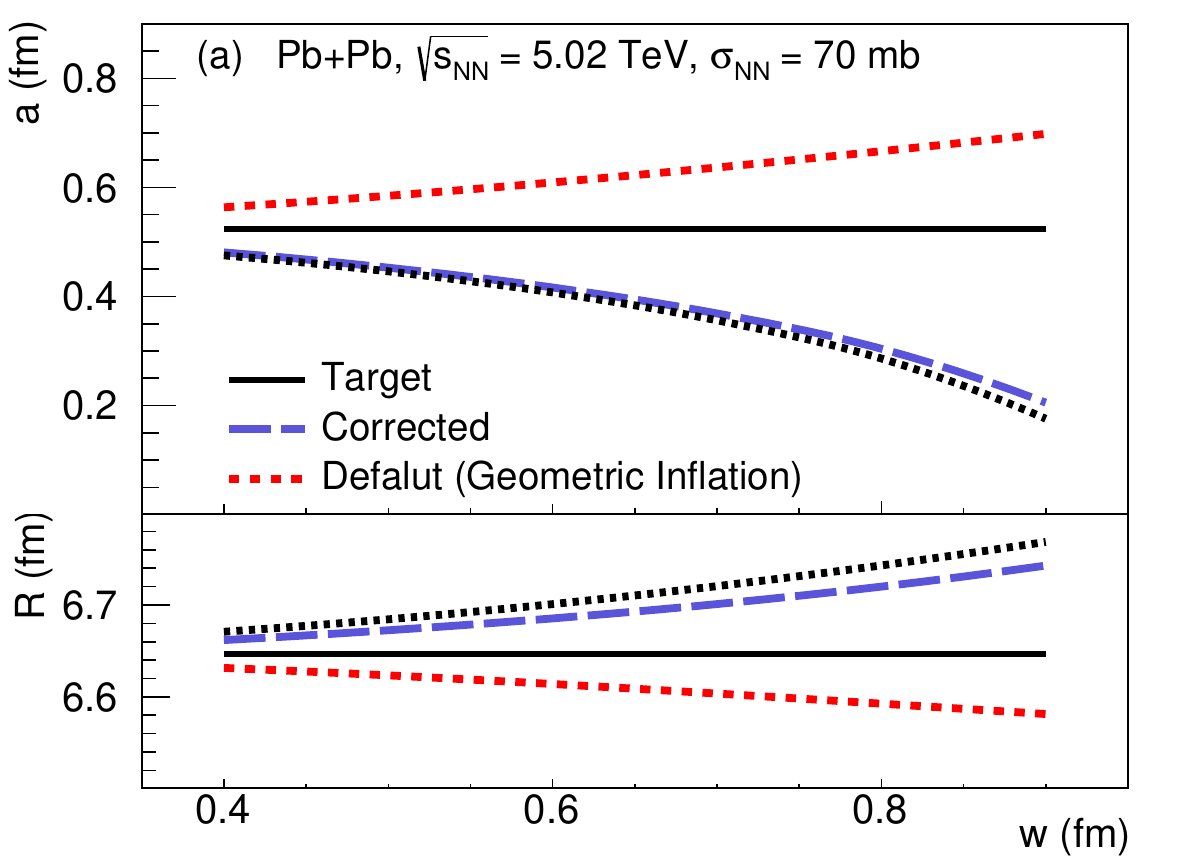}\\
\includegraphics[width=0.49\textwidth]{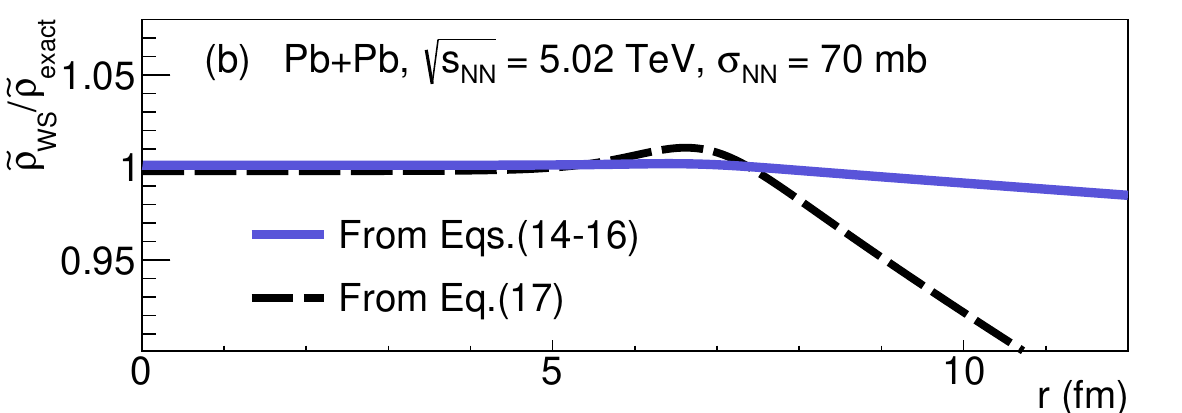}
\caption{Density-consistent Woods--Saxon parameters and validation.  (a) Target parameters $(R,a)$, inflated effective parameters $(R_{\rm eff},a_{\rm eff})$, and corrected input parameters $(\tilde R,\tilde a)$ as functions of $w$.  Dashed curves show the analytical approximation in Eq.~(\ref{eq:ana}).  (b) Comparison between the exact numerical input density obtained from matrix-based deconvolution and the Woods--Saxon profile constrained by the moment-matching method.}
\label{fig:parameter}
\end{figure}

Figure~\ref{fig:parameter}(a) shows how the Woods--Saxon antecedent changes when the \Pb\ density is held fixed.  As $w$ increases, $\tilde a$ decreases to compensate for surface broadening, while $\tilde R$ changes only mildly.  Figure~\ref{fig:parameter}(b) compares the moment-matched Woods--Saxon antecedent with an exact matrix-based deconvolution.  The agreement confirms that the correction implements the fixed-density condition in Eq.~(\ref{eq:TAB_corr}) independently of the Woods--Saxon parametrization.  The propagated difference is numerically negligible: the exact deconvolved density gives $\sigAA=7.411\pm0.002$ b, consistent with $\sigAA=7.406\pm0.002$ b from the moment-matched Woods--Saxon antecedent.

\section{Results}
\begin{figure}[t]
\centering
\includegraphics[width=0.49\textwidth]{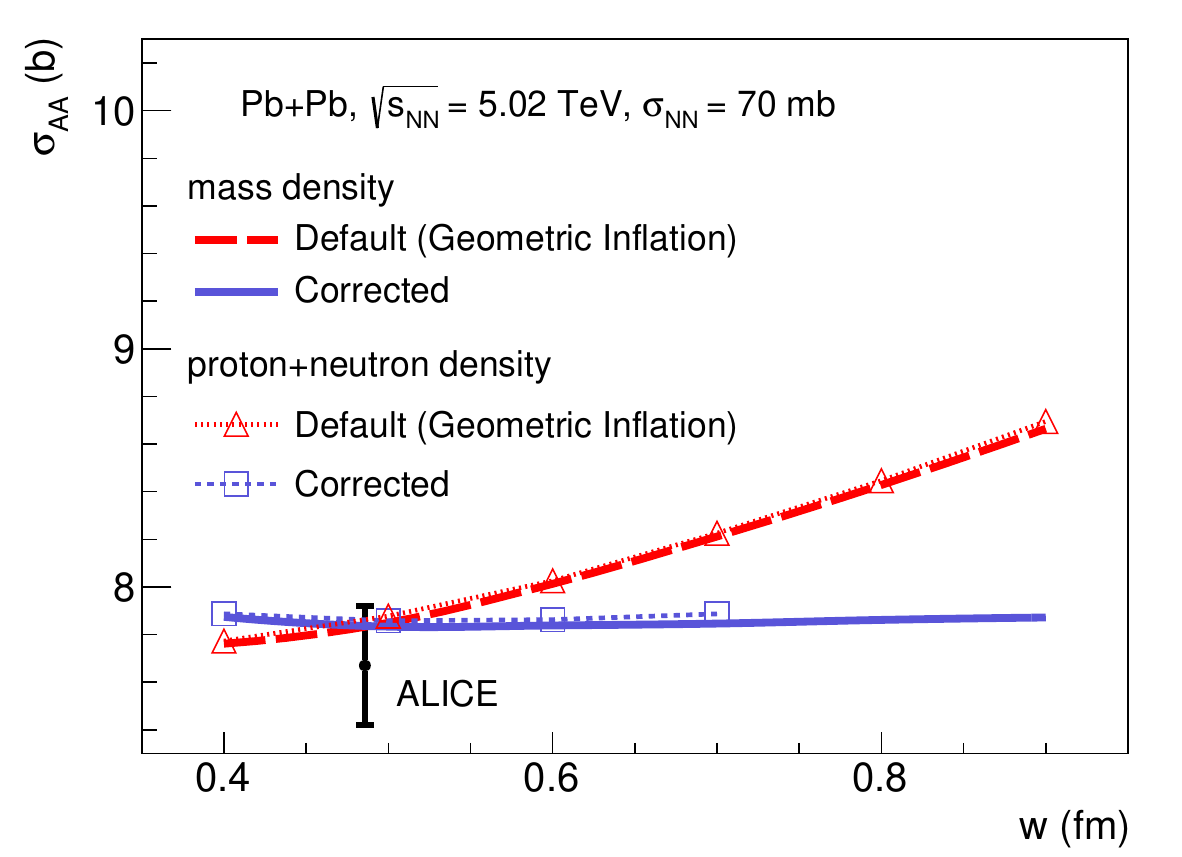}
\caption{Hadronic cross section $\sigAA$ for \PbPb\ collisions at $\snn=5.02$ TeV, computed with the default Woods--Saxon parameters and with Gaussian-kernel-corrected Woods--Saxon parameters, compared with ALICE measurements~\cite{ALICE:2022xir}.  Symbols show the corresponding two-component proton+neutron implementation, which agrees with the single mass-density calculation.}
\label{fig:sigAA}
\end{figure}

Figure~\ref{fig:sigAA} shows the first consequence of the density-preserving baseline.  The same \PbPb\ cross section at $\snn=5.02$ TeV is calculated in two ways: with the default fixed Woods--Saxon center distribution and with the corrected center distribution that preserves the physical mass density.  In the default calculation, increasing $w$ broadens the nuclear surface and produces a strong increase of $\sigAA$.  This is the origin of the strong width sensitivity used in Ref.~\cite{Nijs:2022rme} to constrain the nucleon size.  In the corrected calculation, the physical mass density entering the phase shift is held fixed, and the strong Gaussian-width dependence disappears.

This result gives the main physical message.  The hadronic nuclear cross section is surface sensitive: the interior of a heavy nucleus is already close to black, and the measured interaction probability is controlled by the radial region where the density falls off.  If changing the nucleon width also changes this surface, the cross section changes; if the physical mass density is preserved, the dominant surface effect is removed.  Thus the leading sensitivity of $\sigAA$ is to the physical folded surface supplied to the Glauber calculation.  Information on the finite interaction kernel becomes clean only after this surface has been fixed.  The difference from Ref.~\cite{Nijs:2022rme} is therefore the physical identification of the density being constrained: the same Glauber framework probes the folded nuclear surface when that surface is allowed to vary with the nucleon width.
Once the leading width ambiguity is removed, $\sigAA$ can be used as a surface-sensitive constraint on the nuclear density profile.  Variations of the minimum inter-nucleon distance $d_{\rm min}$ and of $\signn$ within current uncertainties are substantially smaller than the experimental uncertainty, as discussed in~\cite{Nijs:2022rme}.  

To translate the surface constraint into a neutron-skin interval, I use a two-component RMS model.  In a neutron-rich nucleus, the outer surface of the mass density depends on how far the neutron distribution extends beyond the proton distribution.  Quantitatively, this connection follows from
\begin{equation}
R_m^2=R_{\rm p,point}^2+3w^2+\frac{126}{208}(2R_{\rm p,point}\rnp+\rnp^2).
\label{eq:Rms_skin}
\end{equation}
Here $R_{\rm p,point}^2=R_{\rm ch}^2-r_{\rm ch,p}^2$ with $r_{\rm ch,p}=0.84$ fm~\cite{Tiesinga:2021myr}, and $\rnp=\sqrt{R_n^2}-\sqrt{R_p^2}$ is the neutron skin thickness.  At fixed total mass density, a change in the assumed nucleon width can be compensated by a change in $\rnp$.  In the two-component proton+neutron implementation, I use a Gaussian shift method with neutron variance
\begin{equation}
w_n^2=\frac{2R_{\rm p,point}\rnp+\rnp^2}{3}
\label{eq:wn}
\end{equation}
for each sampled neutron, starting from the same profile as the proton density~\cite{Luzum:2023gwy}.

The two-component calculation agrees with the single mass-density calculation, as shown in Fig.~\ref{fig:sigAA}.  The open red symbols show the uncorrected geometric inflation.  When $\rnp$ is adjusted to absorb that inflation and preserve the total mass density, the width independence is restored.  The scan stops when the fixed-density constraint would require a negative neutron skin.

\begin{figure}[t]
\centering
\includegraphics[width=0.49\textwidth]{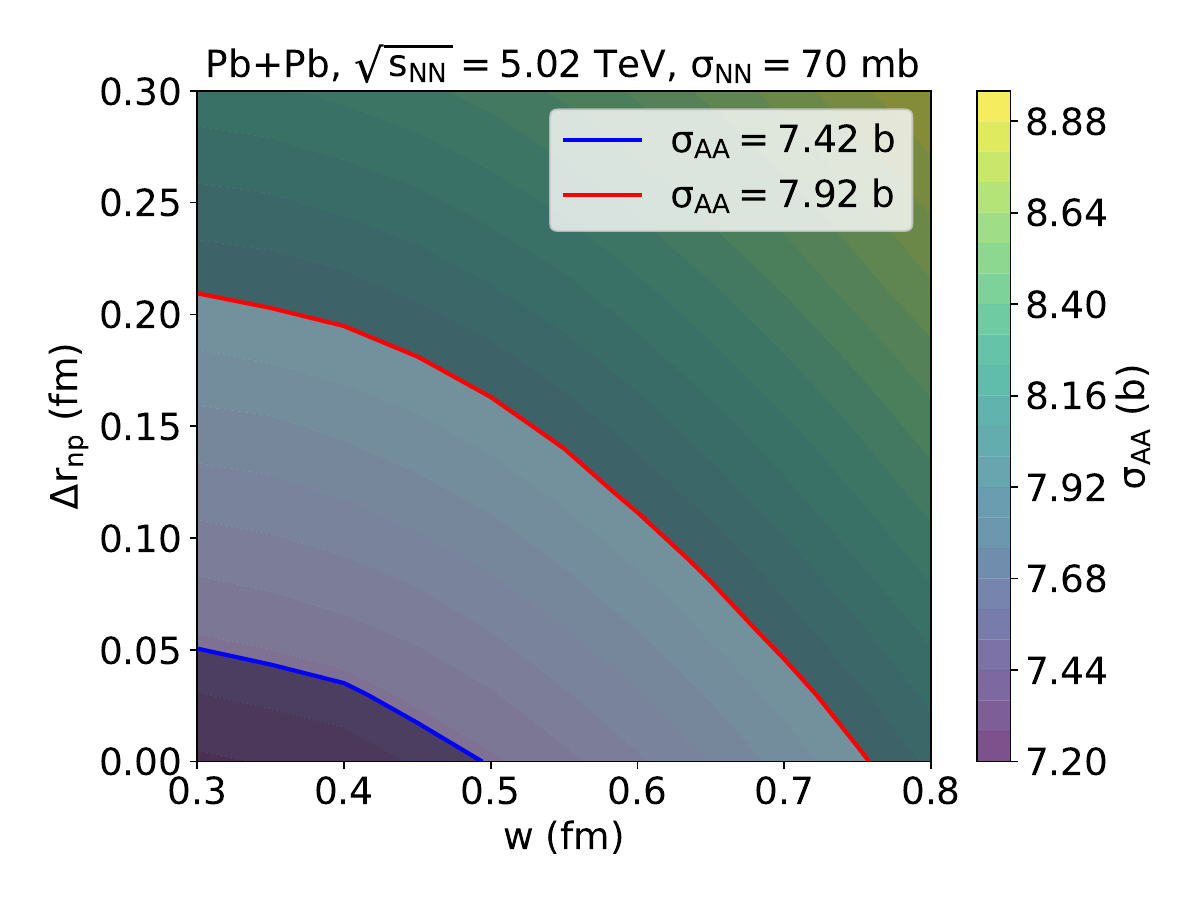}
\caption{Hadronic cross section $\sigAA$ as a function of the nucleon width $w$ and neutron skin thickness $\rnp$ in \PbPb\ collisions at $\snn=5.02$ TeV, compared with the ALICE measurement band~\cite{ALICE:2022xir}.}
\label{fig:neutronskin}
\end{figure}

Figure~\ref{fig:neutronskin} compares $\sigAA(w,\rnp)$ with the ALICE measurement band.  For a reference width $w=0.486$ fm, comparable to the scale associated with the proton charge radius, the data imply $\rnp\in[0,0.17]$ fm.  For a smaller gluonic radius, $w=0.3$ fm~\cite{Caldwell:2010zza}, the neutron distribution must extend further to preserve the mass density, giving $\rnp\in[0.05,0.21]$ fm.  The combined range for $w\in[0.3,0.486]$ fm is therefore
$\rnp\in[0,0.21]\ {\rm fm}$. 
If the empirical relation $\rnp\approx0.101+0.00147L$~\cite{Roca-Maza:2011qcr} is taken merely as a guide, this neutron-skin range still allows for a remarkably wide symmetry-energy slope,
spanning $L\in[-69,74]\ {\rm MeV}$.
This broad interval should be understood as independent high-energy hadronic information on the nuclear surface, complementary to dedicated electroweak and low-energy probes.  Improved measurements of $\sigAA$, together with external constraints on gluonic radii, could sharpen this channel.

These results show that, in a finite-width Glauber model, the nuclear density controlling the global geometry must be distinguished from the finite interaction kernel controlling individual $NN$ interactions.  Hadronic cross sections can be interpreted cleanly only after the nuclear-density baseline is fixed.  With this ambiguity removed, $\sigAA$ is better viewed as a surface-sensitive nuclear-geometry constraint than as a standalone nucleon-size measurement.

\section{Discussion}
The analysis above establishes the density baseline for a Gaussian nucleon profile and clarifies what remains model dependent once this baseline is fixed.  The leading artifact is the change of the folded nuclear mass density when $w$ is varied at fixed point-center density.  After this artifact is removed, residual dependence can still arise from the finite interaction kernel, subnucleonic event-by-event fluctuations, and the nonlinear collision prescription.  These effects represent residual model dependence around the fixed-density baseline.

The Gaussian profile gives the simplest realization of the density-preserving construction.  To estimate the residual model dependence that remains after the density baseline has been fixed, I introduce a one-parameter family of nucleon thickness functions,
\begin{equation}
\tpp(b;\gamma)=\frac{\kappa}{4\pi\zeta^2}\left(\frac{b}{\zeta}\right)^\gamma K_\gamma\left(\frac{b}{\zeta}\right),
\label{eq:thick}
\end{equation}
where $K_\gamma$ is the modified Bessel function of the second kind and $\kappa=2^{1-\gamma}/\Gamma(\gamma+1)$ normalizes the thickness function.  For $\gamma=1$ and $\gamma=3$, Eq.~(\ref{eq:thick}) reproduces monopole and dipole form factors, corresponding to Yukawa-like and exponential nucleon profiles~\cite{Chou:1968bc,Durand:1968ny,Borkowski:1975ume,Wang:1991hta,Bahr:2008pv,dEnterria:2020dwq}.  The Gaussian limit is recovered as $\gamma\to\infty$ with $w=\sqrt{\gamma+1}\zeta$.
The properties of this general nucleon–nucleon thickness function are summarized in Table~\ref{tab:general}.

\begin{table}[t]
\centering
\caption{Properties of the nucleon--nucleon thickness function $\tpp$, single-nucleon form factor, nucleon density, and RMS radius for different shape parameters $\gamma$.  The scale parameters are $\zeta$ for finite $\gamma$ and $w=\zeta\sqrt{\gamma+1}$ for the Gaussian limit.  The constants are $\kappa=2^{1-\gamma}/\Gamma(\gamma+1)$ and $\nu=(1+\gamma)/2$.}
\label{tab:general}
\scriptsize
\resizebox{\textwidth}{!}{%
\begin{tabular}{ccccc}
\toprule
Case & Thickness function $\tpp$ & Form factor $F(q)$ & Nucleon density $\rho(r)$ & RMS radius $r_p$ \\
\midrule
General $\gamma$ &
$\displaystyle \frac{\kappa}{4\pi\zeta^2}\left(\frac{b}{\zeta}\right)^\gamma K_\gamma\left(\frac{b}{\zeta}\right)$ &
$\displaystyle \frac{1}{(1+q^2\zeta^2)^\nu}$ &
$\displaystyle \frac{1}{(2\pi)^{3/2}\zeta^3}\frac{2^{1-\nu}}{\Gamma(\nu)}\left(\frac{r}{\zeta}\right)^{\nu-3/2}K_{\nu-3/2}\left(\frac{r}{\zeta}\right)$ &
$\displaystyle \sqrt{3(1+\gamma)}\,\zeta$ \\
$\gamma=1$ (monopole) &
$\displaystyle \frac{1}{4\pi\zeta^2}\frac{b}{\zeta}K_1\left(\frac{b}{\zeta}\right)$ &
$\displaystyle \frac{1}{1+q^2\zeta^2}$ &
$\displaystyle \frac{1}{4\pi r\zeta^2}e^{-r/\zeta}$ &
$\displaystyle \sqrt{6}\,\zeta$ \\
$\gamma=3$ (dipole) &
$\displaystyle \frac{1}{96\pi\zeta^2}\left(\frac{b}{\zeta}\right)^3K_3\left(\frac{b}{\zeta}\right)$ &
$\displaystyle \frac{1}{(1+q^2\zeta^2)^2}$ &
$\displaystyle \frac{1}{8\pi\zeta^3}e^{-r/\zeta}$ &
$\displaystyle 2\sqrt{3}\,\zeta$ \\
$\gamma\to\infty$ (Gaussian) &
$\displaystyle \frac{1}{4\pi w^2}\exp\left(-\frac{b^2}{4w^2}\right)$ &
$\displaystyle \exp\left(-\frac{q^2w^2}{2}\right)$ &
$\displaystyle \frac{1}{(2\pi)^{3/2}w^3}\exp\left(-\frac{r^2}{2w^2}\right)$ &
$\displaystyle \sqrt{3}\,w$ \\
\bottomrule
\end{tabular}%
}
\end{table}

\begin{figure}[t]
\centering
\includegraphics[width=0.49\textwidth]{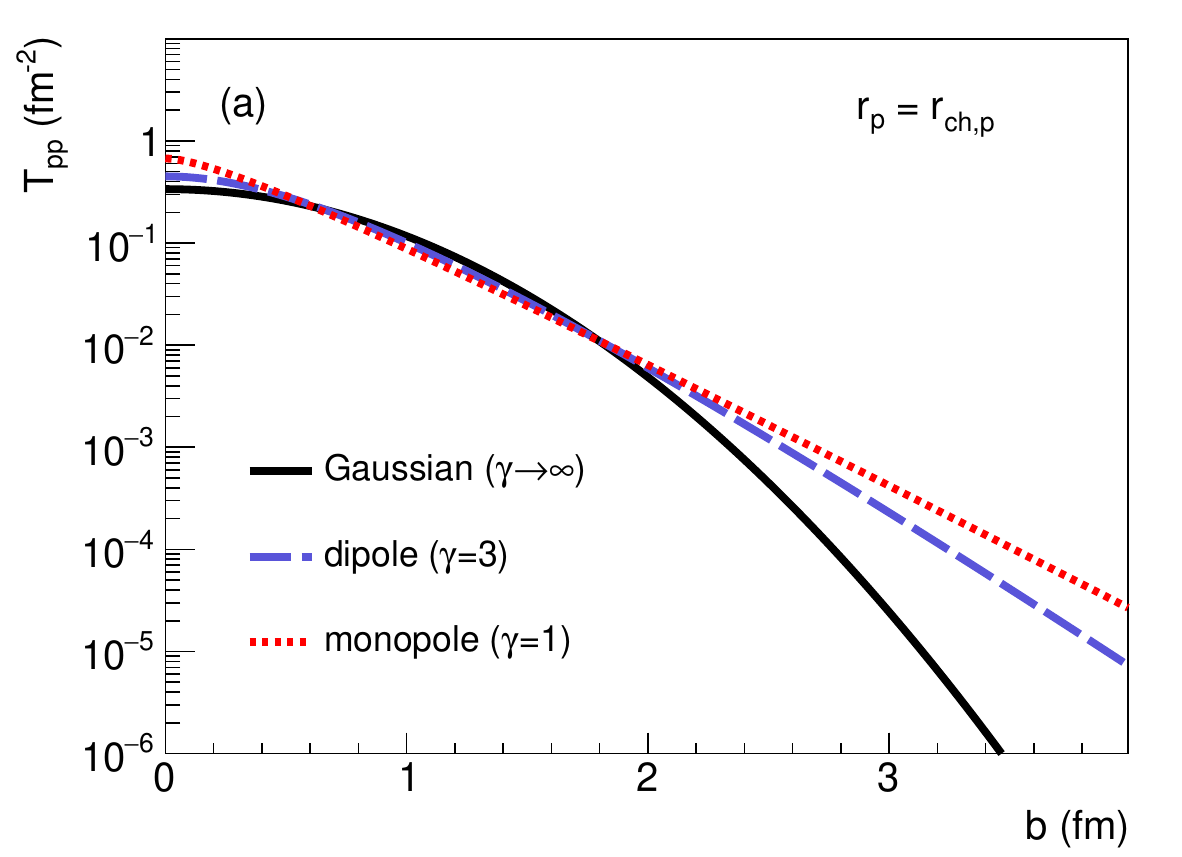}\hfill
\includegraphics[width=0.49\textwidth]{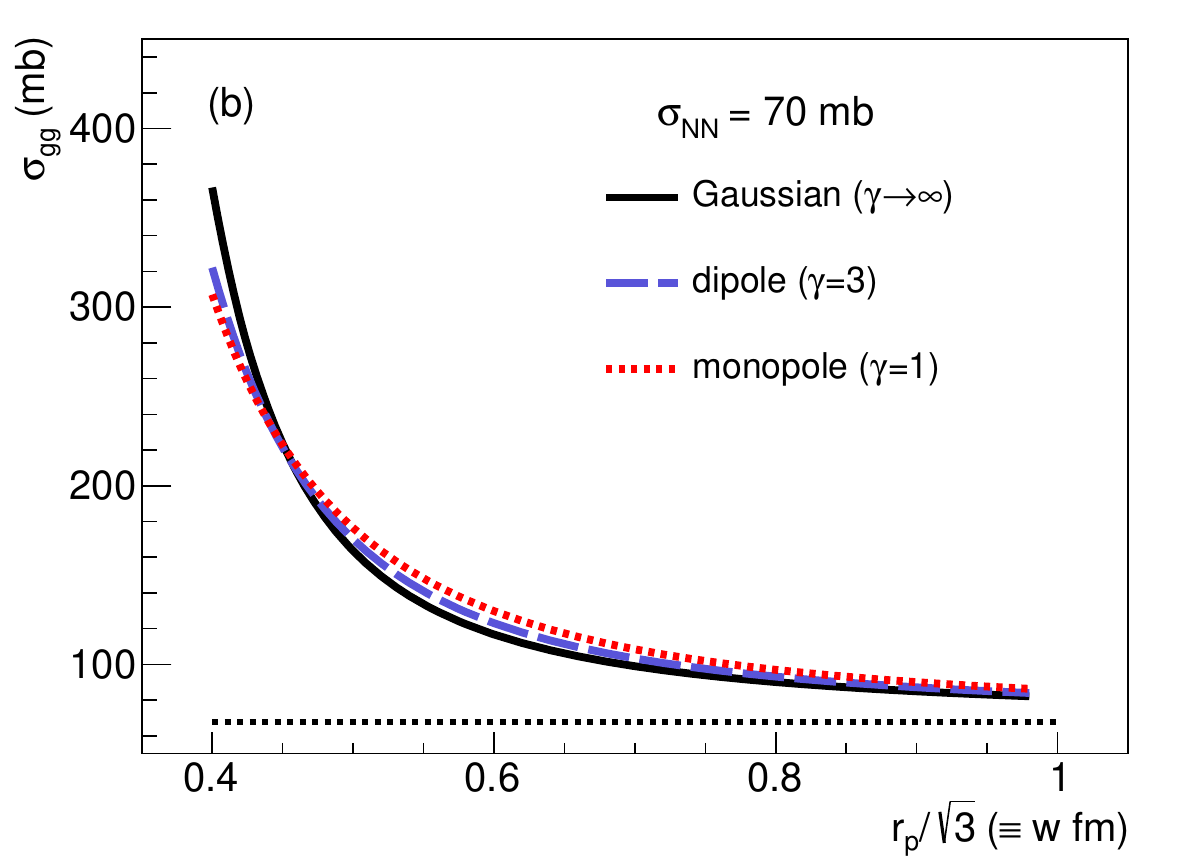}
\caption{Residual kernel dependence.  (a) Nucleon--nucleon thickness function $\tpp$ versus impact parameter for different nucleon form-factor approximations at a common reference RMS radius, chosen here as $r_p=r_{\rm ch,p}$.  (b) Corresponding effective parton--parton cross section $\siggg$ versus nucleon size.  The parameter $\gamma$ is defined by Eq.~(\ref{eq:thick}).}
\label{fig:kernel}
\end{figure}

At this common reference charge RMS radius, the density-corrected input parameters remain close to $\tilde R\simeq6.672$ fm and $\tilde a\simeq0.456$ fm.  The corresponding Monte Carlo cross sections are $\sigAA=7.41$ b for the Gaussian kernel, $7.55$ b for the dipole kernel, and $7.60$ b for the monopole kernel.  This residual variation is different from geometric inflation.  A fixed mass density removes the leading mean-size dependence, while finite-range event granularity, subnucleonic fluctuations, and the nonlinear collision prescription can still produce kernel-dependent corrections.  Gluonic form-factor constraints from $J/\psi$ photoproduction and future electron--ion collider measurements may help reduce this residual uncertainty~\cite{Duran:2022xag,AbdulKhalek:2021gbh}.

Although the present calculation focuses on the hadronic cross section, the same distinction may also affect final-state observables.  In current Bayesian analyses, the nucleon width is often treated as a free parameter and can favor values of order $w\sim1$ fm~\cite{Bernhard:2019bmu,JETSCAPE:2020mzn,Nijs:2020roc}.  Such a width can mix the interaction range, entropy-deposition smearing, and an unintended change of the folded nuclear surface.  A possible direction is to distinguish the width controlling the collision probability from the width controlling the transverse granularity of entropy deposition.  These two widths are physically related but need not be identical.  Keeping this distinction explicit may help reduce the ambiguity in future Bayesian analyses. 

A longer-term implication is that nuclear-structure inputs for heavy-ion collisions may need to become resolution matched: point-proton and point-neutron densities should be accompanied by a clearly specified finite-resolution profile when they are used in high-energy initial-state models.

\section{Conclusions}
This study has revisited a basic question in heavy-ion initial-state modeling: which nuclear density is supplied to the Glauber phase shift when point-nucleon nuclear-structure input is combined with a finite nucleon profile.  In a zero-range Glauber description, the distinction between the sampled center distribution and the density entering the phase shift is immaterial.  Once a finite profile is attached to each sampled center, however, the phase shift depends on the folded density.  Keeping the point-center distribution fixed while varying the nucleon width therefore also changes the folded nuclear surface.

The results show that the strong sensitivity of the hadronic \PbPb\ cross section to the nucleon width is mainly a consequence of this width-induced inflation of the folded nuclear surface.  The density-preserving inverse-convolution construction developed here fixes the density entering the Glauber phase shift by enforcing $\partial_w{\cal T}_{AB}^{\rm corr}\simeq0$.  With this geometric baseline fixed, $\sigAA$ becomes essentially insensitive to a Gaussian nucleon width and instead probes the radial falloff of the nuclear density supplied to the collision model.

This changes the interpretation of hadronic nuclear cross sections.  With the folded nuclear density held fixed, $\sigAA$ is reframed from a nominal nucleon-width constraint into a surface-sensitive nuclear-geometry observable.  For \Pb, the present two-component analysis gives $\rnp\in[0,0.21]$ fm for $w\in[0.3,0.486]$ fm, corresponding to a broad symmetry-energy slope range.  This result illustrates how high-energy hadronic information can be connected to conventional nuclear-structure variables once the density baseline is specified.

The broader implication is that nuclear geometry cannot be transferred from low-energy structure theory to high-energy collision simulations without specifying the resolution at which that geometry is probed.  Resolution matching therefore provides a necessary baseline for future quantitative connections among nuclear structure, hadronic cross sections, and QGP initial conditions.

\section*{Acknowledgements}
The author thanks Dandan Zhang, Yingxun Zhang, and Qing Zhao for interesting discussions.  This work is supported in part by the National Natural Science Foundation of China under Grant No. 12275082.

\bibliography{ref}

\end{document}